\documentclass[a4paper,10pt]{article}
\usepackage[utf8]{inputenc}
\usepackage{graphicx}
\usepackage{txfonts}
\usepackage{natbib}

\newcommand{\teff}  {T$_\mathrm{eff}$}
\newcommand{\logg}  {$\log g$}

\title{Chemical abundances of neutron capture elements in exoplanet-hosting stars\thanks{
Based on observations collected at the La Silla Observatory, ESO
(Chile), with the HARPS spectrograph at the 3.6 m ESO telescope (ESO
runs ID 72.C---0488, 082.C---0212, and 085.C---0063).}
}

\author{E. Delgado Mena,$^{1}$\thanks{E-mail: \href{mailto:elisa.delgado@astro.up.pt}{elisa.delgado@astro.up.pt}}
V. Zh. Adibekyan,$^{1}$
P.~Figueira,$^{1,2}$\\
J.~I.~Gonz\'alez Hern\'andez,$^{3,4}$
N.~C.~Santos,$^{1,5}$
M. Tsantaki,$^{6}$\\
S. G. Sousa,$^{1}$
J.P.~Faria,$^{1}$
L. Su\'arez-Andr\'es,$^{3,4}$ 
and G. Israelian$^{3,4}$\\     
\scriptsize
$^{1}$Instituto de Astrof\'isica e Ci\^encias do Espa\c{c}o, Universidade do Porto, CAUP, Rua das Estrelas, PT4150-762 Porto, Portugal\\
\scriptsize
$^{2}$European Southern Observatory, Alonso de Cordova 3107, Vitacura, Santiago, Chile\\
\scriptsize
$^{3}$Instituto de Astrof\'{\i}sica de Canarias, C/ Via Lactea, s/n, 38205, La Laguna, Tenerife, Spain \\
\scriptsize
$^{4}$Departamento de Astrof\'isica, Universidad de La Laguna, 38206 La Laguna, Tenerife, Spain\\
\scriptsize
$^{5}$Departamento de F\'isica e Astronom\'ia, Faculdade de Ci\^encias, Universidade do Porto, Portugal\\
\scriptsize
$^{6}$Instituto de Radioastronom\'ia y Astrof\'isica, IRyA, UNAM, Campus Morelia, A.P. 3-72, C.P. 58089, Michoac\'an, Mexico\\
}

\begin{document}

\maketitle

\begin{abstract}
To understand the formation and composition of planetary systems it is important to study their host stars composition since both are formed in the same stellar nebula. In this work we analyze the behaviour of chemical abundances of Cu, Zn, Sr, Y, Zr, Ba, Ce, Nd and Eu in the large and homogeneous HARPS-GTO planet search sample ($R \sim$\,115000). This sample is composed of 120 stars hosting high-mass planets, 29 stars hosting exclusively Neptunians and Super-Earths and 910 stars without detected giant planets. We compare the [X/Fe] ratios of such elements in different metallicity bins and we find that planet hosts present higher abundances of Zn for [Fe/H]$<$--0.1\,dex. On the other hand, Ba, Sr, Ce and Zr abundances are underabundant in stars with planets, with a bigger difference for stars only hosting low-mass planets. However, most of the offsets found can be explained by differences in stellar parameters and by the fact that planet hosts at low metallicity mostly belong to the Galactic thick disk. Only in the case of Ba we find a statistically significant (3$\sigma$) underabundance of 0.03\,dex for low-mass planet hosts. The origin of these elements is quite complex due to their evolution during the history of the Galaxy. Therefore, it is necessary to understand and characterize the stellar populations to which planet hosts belong in order to do a fair comparison with stars without detected planets. This work demonstrates that the effects of Galactic chemical evolution and not the presence of planets mostly account for the differences we find.

\end{abstract}

\section{Introduction}

In the last 20 years the race to search for planets with the radial velocity ($RV$) method has provided large sets of high quality and high-resolution spectra, allowing the scientific community to explore the chemical abundances of stars with and without planets and search for possible differences among them. After the first discoveries of exoplanets it was soon proposed that stars hosting close-in giant planets present metallicities (defined as iron abundance) higher than stars with no detected planets \citep{gonzalez97,santos01}. This finding was further confirmed with larger samples of planet hosts stars \citep[e.g.][]{santos04,fischer05,sousa08} establishing the currently well-known giant planet metallicity\footnote{We note that through the rest of the paper we will use in an indistinguishable way the terms metallicity or [Fe/H] to both refer to the abundance of iron respect to the Sun} correlation. Interestingly, a recent work by \citet{santos17} shows that stars hosting planets more massive than 4\,M$_{J}$ tend to be more metal-poor and more massive than stars hosting planets with mass below this threshold. On the other hand, less massive planets than 4\,M$_{J}$ are preferentially found around metal-rich stars, following the aforementioned planet-metallicity correlation. This finding might indicate that two different processes of planet formation are at play.

On the other hand, more recent studies have shown that this correlation does not seem to hold for stars hosting less massive planets, those with masses as Neptune or lower \citep{sousa_harps4,buchhave12,everett13,buchhave15}. However, other works do find that small planet formation is more probable around metal rich stars than around metal poor stars though with a lower factor than giant planet formation \citep{wang15}. Moreover, the work by \citet{zhu16} claims that the high occurrence rate for low-mass planets and the current small size samples of low-mass planet hosts do not allow to detect the planet metallicity correlation which is however recovered by their theoretical simulations. 

The giant planet metallicity correlation supports the core-accretion scenario for the formation of planets \citep{pollack96,ida04,mordasini09} in which it is assumed that planetesimals are formed by the condensation of heavy elements. Also, it seconds \citet{nayakshin15} theory of Tidal downsizing. These works show how important is the study of stellar metallicities of planet hosts to constrain the planet formation theories. Although most studies of abundances in planet host stars have traditionally used iron as a metallicity proxy, detailed abundance analysis have also allowed to reach a number of interesting conclusions. Some early works suggested possible enrichment in some species such as Ca, Na, Si, Ni, Ti, V, Co, Mg, Al or Mn \citep{gonzalez01,santos00,sadakane02,bodaghee03,fischer05,beirao05,gilli06,bond06,robinson06,gonzalez07,kang11,brugamyer11}. However, the results were not conclusive or they were based on very small samples in some cases. On the other hand, other studies reported no differences in the abundance ratios [X/Fe] among stars with and without planets \citep{takeda07,bond08,neves09,jonay10,suarez-andres16}.

\begin{figure*}
\centering
\includegraphics[width=1.0\linewidth]{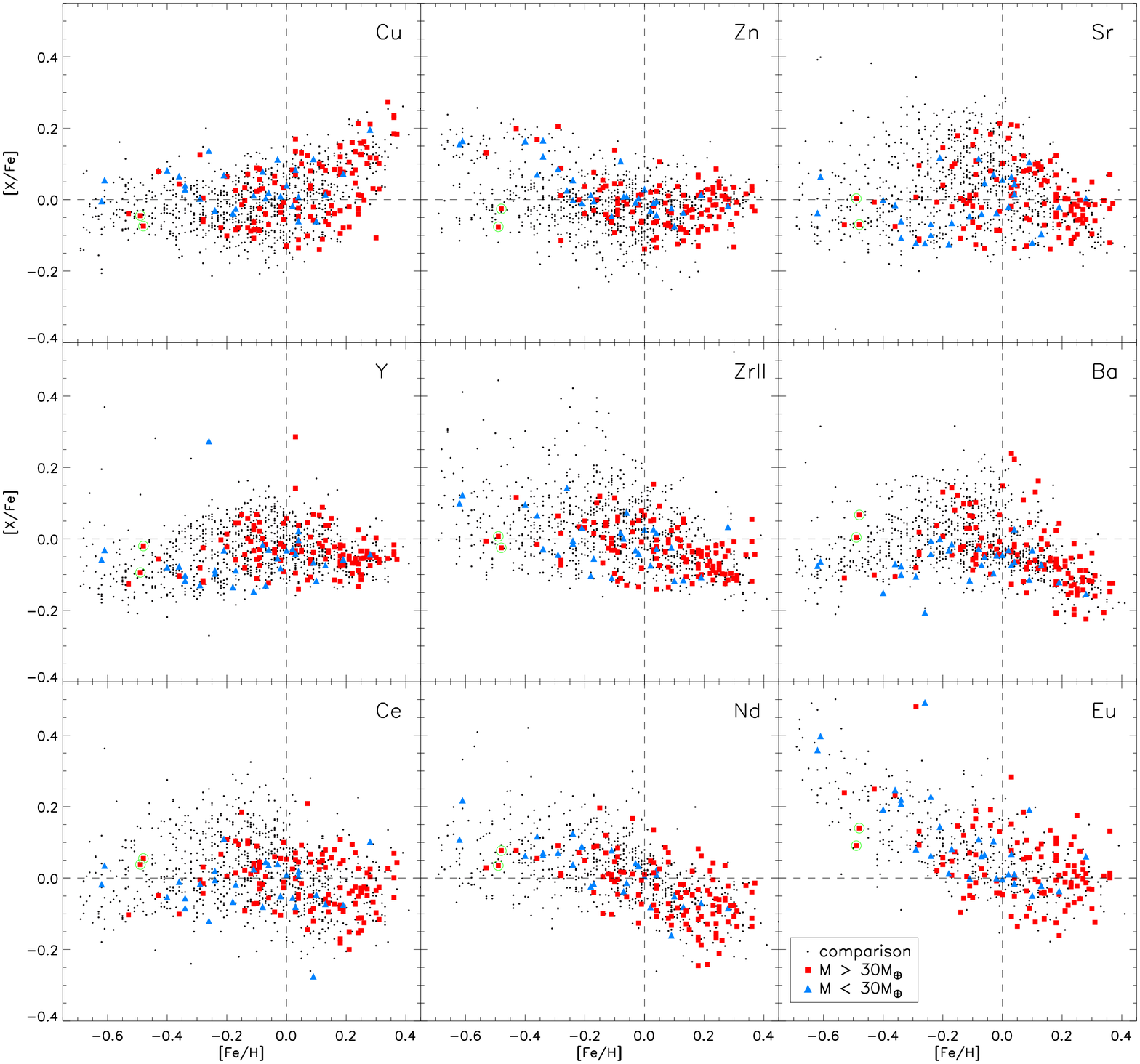}
\caption{Final [X/Fe] ratios as a function of [Fe/H] for stars in or sample. Low mass planets are depicted with blue triangles meanwhile stars with high-mass planets are shown with red squares. Black dots represent single stars. The two Jupiter-like hosts with a circle around a square are HD 171028 and HD 190984.} 
\label{all_XFe_Fe}
\end{figure*}

\begin{figure}
\centering
\includegraphics[width=1.0\linewidth]{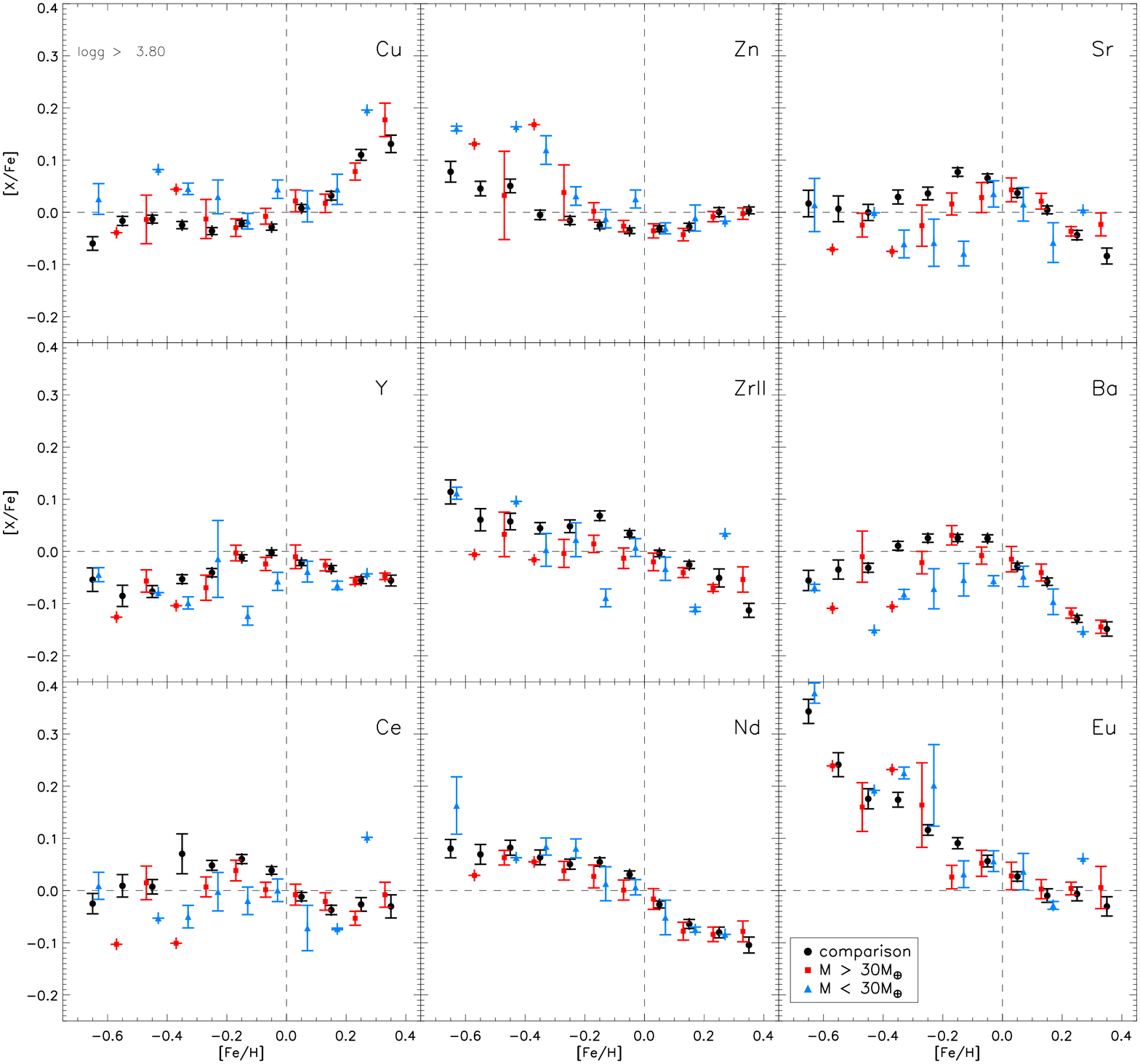}
\caption{Final [X/Fe] ratios as a function of [Fe/H] for all the stars in our sample. The mean abundances in each metallicity bin of 0.1\,dex are shown together with the standard error of the mean. Symbols as in Fig. \ref{all_XFe_Fe}. The symbols for planet hosts are slightly shifted in [Fe/H] position inside each bin for the sake of clarity.} 
\label{bins_XFe_Fe}
\end{figure}

In a previous work, using the HARPS-GTO sample composed of 1111 stars, \citet{adibekyan12_alpha} found that the [X/Fe] ratios of Mg, Al, Si, Sc, and Ti both for giant and low-mass planet hosts are systematically higher than those of stars without detected planets at low metallicities ([Fe/H]\,$\lesssim$ from --0.2 to 0.1\,dex depending on the element). Moreover, this work confirmed that planets form preferentially in the thick disk rather than in the thin disk at lower metallicities. The reason for this preference is probably that if the amount of iron is low it needs to be compensated with other elements important for planet formation, such as Mg and Si \citep[e.g.][]{bond10} and these elements are more abundant in the thick disk. This result was further confirmed using \textit{Kepler} targets with small planets \citep{adibekyan12_kepler}. The earlier works by \citet{haywood08,haywood09} also found that giant planet hosts at lower metallicities belong to the thick disk. However, these authors suggest that a parameter related to the galactocentric radius (in particular, the molecular hydrogen) and not the metallicity might be responsible for the formation of planets. 

The aim of this study is to extend our previous works in this field towards heavier elements, since they have not been as extensively studied in the literature as light, iron-peak or $\alpha$ elements. The first works based on small samples could not find any differences for Ba \citep{huang05} or Zn and Eu \citep{gonzalez07}. \citet{bond08} was the first study focused on deriving \textit{r-} and \textit{s-}process elements in exoplanet hosts stars but they did not detect any dissimilarity between stars with and without planets for Y, Zr, Ba, Nd and Eu. Recently, \citet{dasilva15} and \citet{jofre_planets15} have found differences in Ba abundances for dwarf and evolved stars with and without planets. Finally, the work by \cite{mishenina16}, who analyze several heavy elements, also reports an underabundance of Ba for exoplanet hosting stars. 

Our large and homogeneous sample of stars with and without detected planets gives us the opportunity to study possible correlations between planet occurrence and heavy element abundances in a very detailed way. This paper is organized as follows: in Sect. 2 we briefly introduce the sample of stars and abundances used in this study. The results are discussed in Sect. 3. Finally, in Sect. 4, we draw our main conclusions.

\section{Data}

The baseline sample used in this work consists of 1111 FGK stars observed within the context of the HARPS GTO planet search programs \citep{mayor03,locurto,santos_harps4}. The final spectra have a resolution of R $\sim$115000 and high signal-to-noise ratio (45\% of the spectra have 100\,$<$\,S/N\,$<$\,300, 40\% of the spectra have S/N\,$>$\,300 and the mean S/N is 380). The total sample is composed by 151 stars with planets and 960 stars without detected planets (hereafter single stars\footnote{We note that due to the precision of radial velocities achieved with HARPS and the long-term survey we can be sure that giant planets do not exist around such stars but we cannot rule out the presence of small planets, especially at long periods.} although the final sample used here is slightly smaller as explained below.

Precise stellar parameters were homogeneously re-derived in \citet[][hereafter DM17]{delgado17} following the method of \cite{sousa08} and applying a special linelist for cool stars \cite{tsantaki13}. We also corrected the spectroscopic gravities. This sample with corrected stellar parameters is composed of 1059 stars (29 stars with Neptunian mass planets, 120 stars hosting Jupiter-like planets and 910 single stars). Our stars have typical effective temperatures (\teff) values between 4500\,K and 6500\,K and surface gravities (\logg) mostly lie in the range 4\,$<$\,$\log g$\,$<$\,5 dex meanwhile the metallicity covers the range -1.39\,$<$\,[Fe/H]\,$<$\,0.55\,dex. Chemical abundances of Cu, Zn, Sr, Y, Zr, Ba, Ce, Nd and Eu were determined under local thermodynamic equilibrium (LTE) using the 2014 version of the code MOOG \citep{sneden} and a grid of Kurucz ATLAS9 atmospheres \citep{kurucz}. For more details about the sample and the analysis we refer the reader to DM17. Chemical abundances of these sample stars for refractory elements with atomic number $<$ 29 can be found in \citet{adibekyan12} together with oxygen \citep{bertrandelis15}, carbon \citep{suarez-andres17}, lithium \citep{delgado14,delgado15} and beryllium and nitrogen abundances \citep[][only for a subsample of stars]{santos_be_plan,delgado_Be2,suarez-andres16}.

\begin{figure}
\centering
\includegraphics[width=7cm]{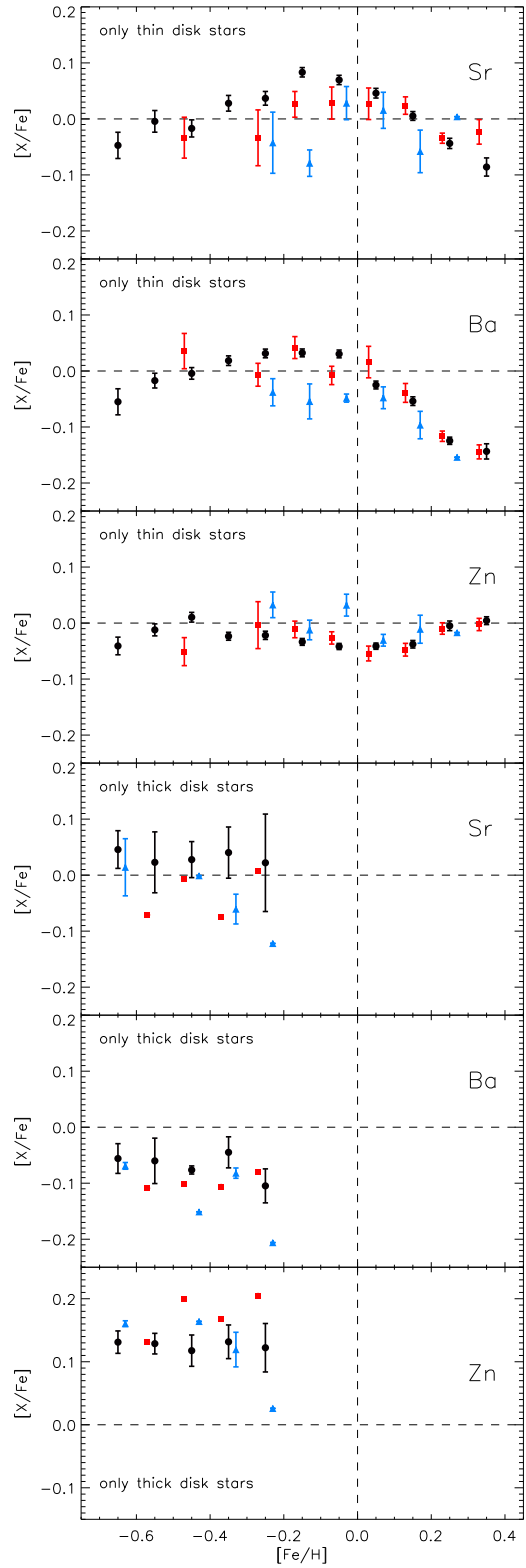}
\caption{Final [Sr/Fe], [Ba/Fe], [Zn/Fe] ratios as a function of [Fe/H] for the separated populations of thin (upper panels) and thick disk (lower panels). The mean abundances in each metallicity bin of 0.1\,dex are shown together with the standard error of the mean. Symbols as in Fig. \ref{all_XFe_Fe}.} 
\label{bins_XFe_Fe_thin}
\end{figure}

\section{Results and discussion} \label{sec:results}

A traditional way of evaluating the differences between planet hosts and single stars is to compare their abundance distributions. Earlier works showed that the [X/H] ratios of planet hosts are shifted towards higher values as compared to single stars \cite[e.g.][]{ecuvillon06,bond08,neves09}. This can be just a consequence of the Galactic chemical evolution (GCE), since generally, the abundances ratios [X/H] increase with [Fe/H] and there is a higher probability of forming giant planets in chemical enriched clouds \citep[e.g.][]{santos04}. However, it is more interesting to compare the [X/Fe] ratios at similar metallicities to diminish the effects of GCE. In Fig. \ref{all_XFe_Fe} we show the [X/Fe] ratios for all the stars in our sample. We have divided the planet hosts in two groups, depending on the mass of the heaviest planet in the system\footnote{The planet masses are extracted from The Extrasolar
Planets Encyclopaedia, exoplanet.eu}, following the work by \citet{adibekyan12_alpha}: those hosting only Neptunians or Super-Earths (M\,$<$\,30\,M$_{\oplus}$, blue triangles) and those with high-mass companions (M\,$\geq$\,30\,M$_{\oplus}$, red squares). 

From a first look it seems that at super-solar metallicities the planets hosts are well mixed with the single stars. However, for lower metallicities it is clear that stars with planets show abundances of Zn higher on average than single stars meanwhile Ba, Ce and Sr seem to be underabundant at [Fe/H]\,$\lesssim$\--0.1\,dex for planet hosts. In order to better appreciate these differences we show in Fig. \ref{bins_XFe_Fe} the average [X/Fe] ratios in 0.1\,dex metallicity bins together with the standard error of the mean. Having such small [Fe/H] bins allows us to reduce any GCE effect although one must be careful when comparing stars from different populations (thin and thick disk) at the same [Fe/H] because the GCE influences the abundances of those elements in a different way \citep[e.g.][DM17]{mishenina13,mikolaitis17,battistini16}. In our work the separation between the thin and the thick disk is made with a chemical criteria, using the abundances of $\alpha$ elements. We refer the reader to \cite{adibekyan11} and DM17 for more details. 

\begin{figure}
\centering
\includegraphics[width=17cm]{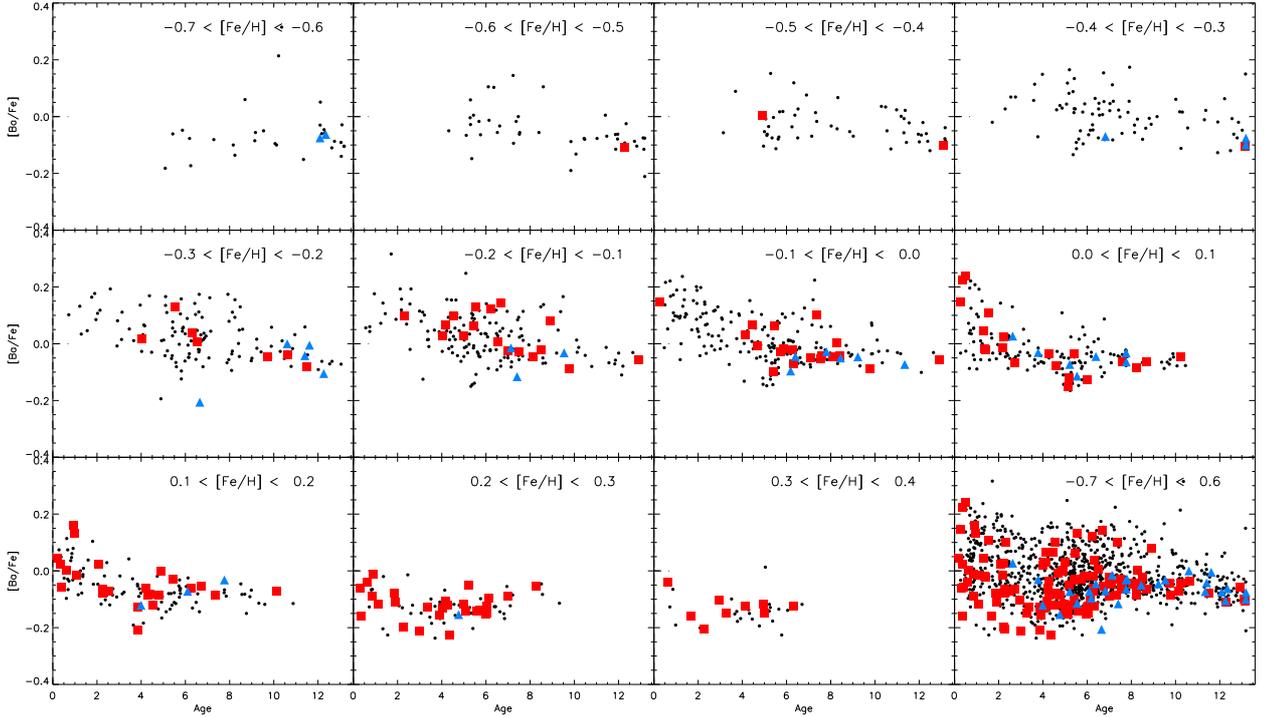}
\caption{[Ba/Fe] ratios as a function of age in all the sample (bottom right panel) and in different [Fe/H] bins. Symbols as in Fig. \ref{all_XFe_Fe}.} 
\label{Ba_age}
\end{figure}

\subsection{Overabundance of Zn, Cu and Eu in planet hosts}
In Fig. \ref{bins_XFe_Fe} we can see how the average abundance of Zn for planet hosts starts to rise at [Fe/H]\,$\lesssim$\,--0.2\,dex towards low metallicities compared to single stars, reaching values by $\sim$0.15\,dex higher in some bins. This overabundance takes place in stars with either light or massive planets. The higher abundances of Zn resembles the overabundance of $\alpha$ elements in stars hosting exoplanets discovered by \citet{adibekyan12_alpha}. Only in the bin --0.5\,$<$\,[Fe/H]\,$<$\,--0.4\,dex, the stars with massive planets (red squares) have a similar average Zn as single stars. This is caused by the two slightly evolved stars HD 171028 and HD 190984 (\logg\ values of 3.83\,dex and 3.92\,dex, respectively) which belong to the thin disk and do not present enhanced $\alpha$ abundances nor enhanced Zn. For the sake of clarity, these stars are depicted with circled red squares in Fig. \ref{all_XFe_Fe} that is very similar to Fig. 5 of \citet{adibekyan12_alpha}. 

It has been shown that there is a clear separation of Zn abundances between the thin and the thick disk \citep[][ DM17]{mikolaitis17}. Therefore, Zn abundances behave in a similar way to $\alpha$ abundances since half of the Zn is produced by $\alpha$ freezout in neutrino winds during supernova explosions of massive stars \citep[e.g.][and refererences therein]{bisterzo05}. As a consequence, it is not surprising that [Zn/Fe] ratios in planet hosts at low [Fe/H] are also enhanced since most of our stars with planets at low [Fe/H] do belong to the thick disk as shown by \citet{adibekyan12_alpha}. In detail, all the planet hosts in our sample with [Fe/H]\,$<$\,--0.3\,dex belong to the thick disk, expect the two slightly evolved stars. Indeed, if we compare the stars of the thin or thick disk separately (see Fig. \ref{bins_XFe_Fe_thin}) the differences would be negligible in most of the bins. 
Theoretical studies have shown that the bulk composition of terrestrial planets is made of Fe, Mg, Si and O \citep[e.g.][]{bond10,elser12,thiabaud14,dorn15}, thus, a priori we do not expect that this Zn enhancement is related to terrestrial planet formation but just a consequence of Zn having a similar GCE as Mg and Si. On the other hand, the composition of the cores in giant planets is not so well understood so we cannot discard that this kind of element may have an implication for giant planet formation. Moreover, in planet formation models the dust-to-gas ratio is a clue factor, which is proportional to metallicity, and thus the content of all the metals is important. Nevertheless, we note that the solar abundance of Zn relative to other major building-block elements is quite small.

We note that although the abundances of Eu behave in a similar way to $\alpha$ elements as shown in DM17, only in some [Fe/H] bins we can find higher abundances for planet hosts but the larger errors of Eu abundances prevent us from confirming any clear difference. In Fig. \ref{bins_XFe_Fe} we can also see that [Cu/Fe] is slightly overabundant in planet hosts at [Fe/H]\,$<$\,--0.2\,dex. Again, this can be due to the fact of planets belonging to the thick disk since in DM17 (see their Fig. 10) we show how Cu is systematically higher in the thick disk when compared to their thin disk counterparts (though the difference is lower than in the case of Zn). 

\begin{figure}
\centering
\includegraphics[width=17cm]{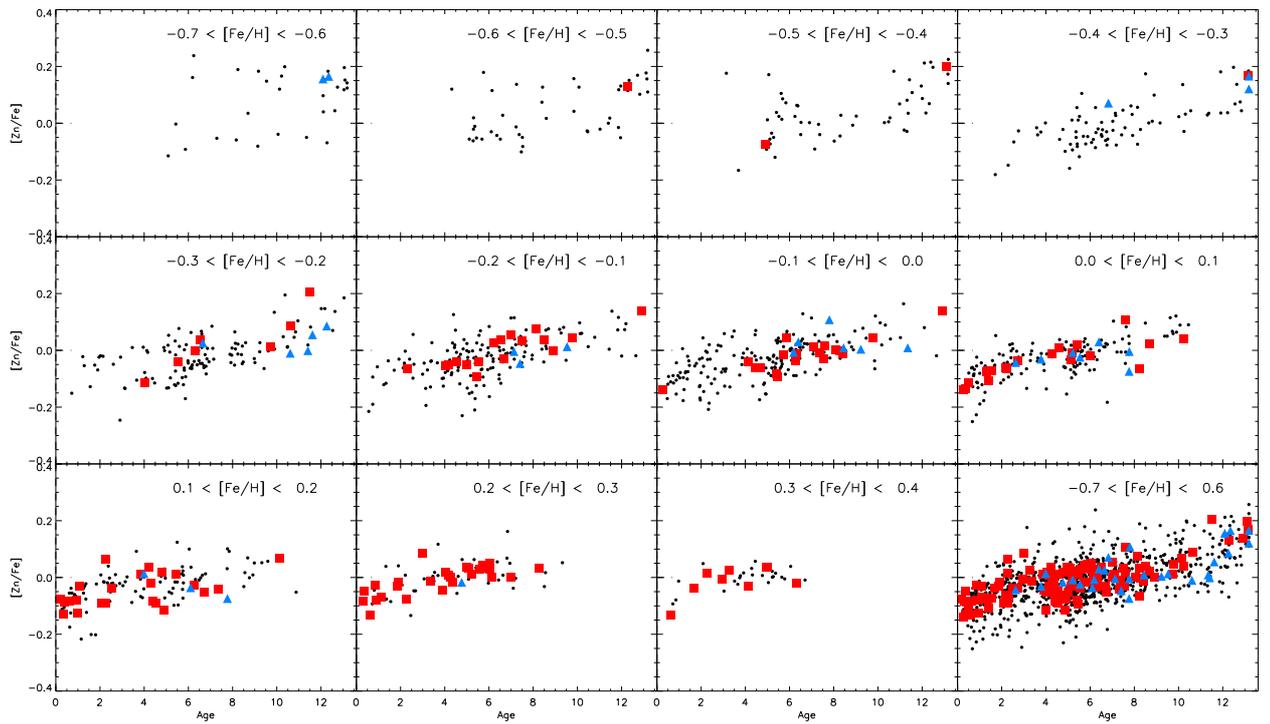}
\caption{[Zn/Fe] ratios as a function of age in all the sample (bottom right panel) and in different [Fe/H] bins. Symbols as in Fig. \ref{all_XFe_Fe}.} 
\label{Zn_age}
\end{figure}

\begin{figure}
\centering
\includegraphics[width=17cm]{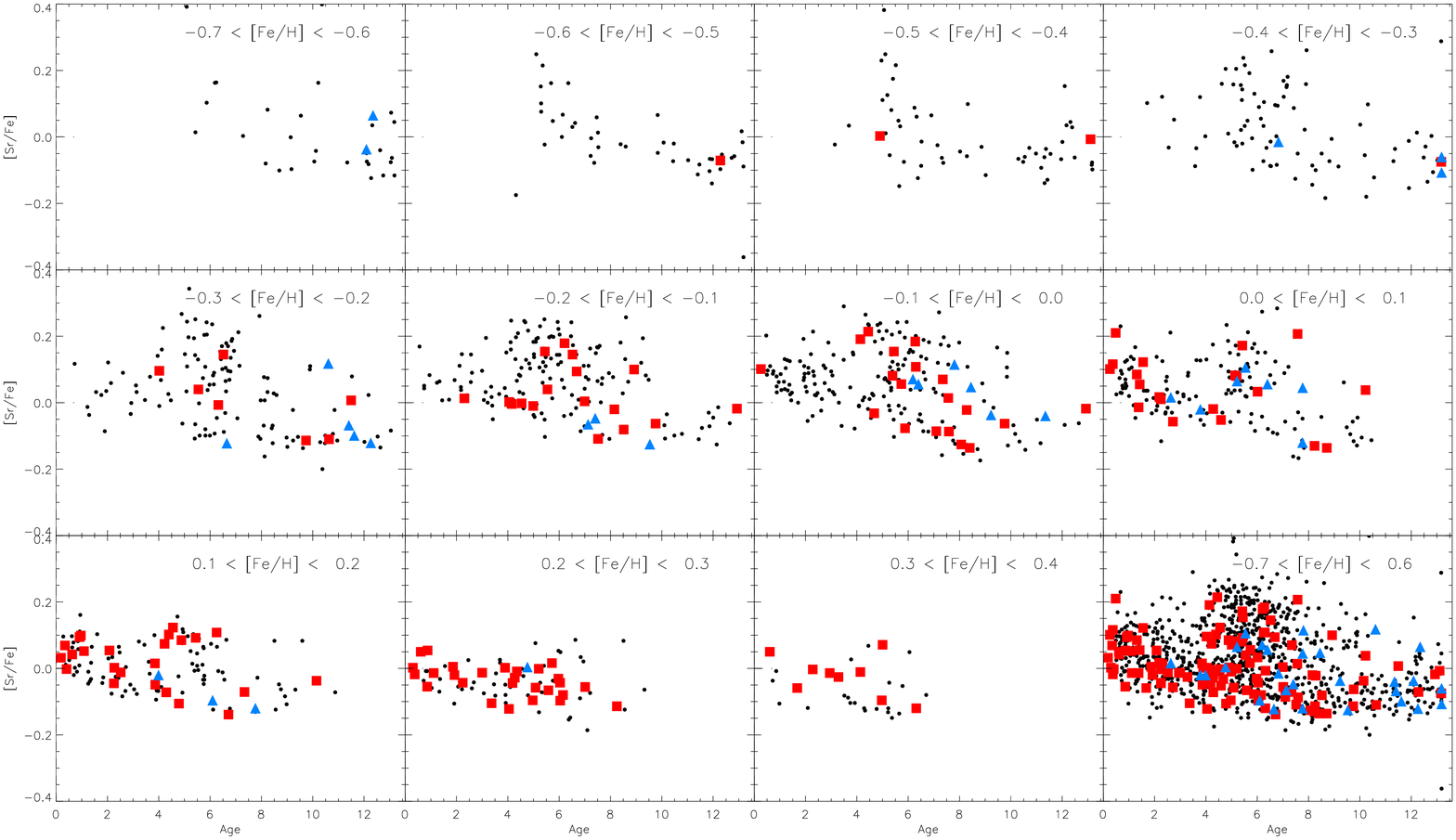}
\caption{[Sr/Fe] ratios as a function of age in all the sample (bottom right panel) and in different [Fe/H] bins. Symbols as in Fig. \ref{all_XFe_Fe}.} 
\label{Sr_age}
\end{figure}

\subsection{Underabundance of Ba, Sr, Ce, and Zr in planet hosts}
The panel for Ba in Fig. \ref{bins_XFe_Fe} shows that this element is depleted in planet hosts in most of the metallicity bins. For [Fe/H]\,$<$\,0\,dex low-mass planet hosts present systematically lower abundances of Ba than single stars. On the other hand, stars with massive planets show lower abundances as well, but the difference is only significant in some bins. Ba abundances in the thick disk are lower on average than in the thin disk, especially in the range --0.5\,$<$\,[Fe/H]\,$<$\,--0.2\,dex. Since many of our low-mass planet hosts belong to the thick disk this could be the reason of the underabundance. However, if we separate the thin and thick disk populations (see Fig. \ref{bins_XFe_Fe_thin}) we still observe the depletion of Ba starting at [Fe/H]\,$<$\,0\,dex for thin disk Neptunian hosts and for all the thick disk planet hosts except in the most metal-poor bin. We note that by dividing the population of thick/thin disk stars we reduce the number of planets hosts in each sub-group. At --0.7\,$<$\,[Fe/H]\,$<$\,--0.2\,dex we have 69 single stars, 5 high-mass and 7 low-mass planet hosts in the thick disk, meanwhile in the thin disk we have 233, 8 and 4 objects respectively. We remark that the population of high-$\alpha$ metal-rich stars \citep[discovered by][]{adibekyan11} presents Ba abundances lower on average than thin disk stars (DM17) but only $\sim$\,10\,\% of our planet hosts belong to that population, thus it cannot be the reason for the apparent Ba depletion in stars with planets at [Fe/H]\,$>$\,--0.2\,dex.

The case for Sr is similar to Ba. In the --0.6\,$<$\,[Fe/H]\,$<$\,0.0\,dex range, high-mass planets hosts show a moderate depletion with respect to single stars  meanwhile low-mass planet hosts show the greatest differences, especially in the region --0.4\,$<$\,[Fe/H]\,$<$\,--0.1\,dex. If we compare the abundances among only the thin disk or thick disk stars the difference remains (see Fig. \ref{bins_XFe_Fe_thin}), and we can find underabundances of more than 0.1\,dex in low-mass planet hosts. The populations of thin and thick disk present well mixed abundances of Sr in the range --0.6\,$<$\,[Fe/H]\,$<$\,--0.2\,dex (see Fig. 10 of DM17). Therefore, the underabundance of Sr in planet hosts cannot be explained only by the fact that they mostly belong to the thick disk. 
 
In Figs. \ref{all_XFe_Fe} and \ref{bins_XFe_Fe} we can also observe a similar pattern of depletion for Ce abundances in planet hosts. This can be expected since Ba and Ce are both heavy-\textit{s} elements with a similar percentage of contribution from the \textit{s}-process, hence a similar GCE. A similar level of underabundance is found for Zr (a light-\textit{s} element as Sr), more systematic for massive planet hosts. However, we note that the underabundance of Zr and Ce seems lower than for Sr and Ba and that errors in Zr and Ce abundances are higher, thus this offset should be taken with caution.

Several works already noted the depletion of Ba in planet host stars. \citet{dasilva15} found that giant planet hosts around dwarf stars in the thin disk are not enriched in [Ba/H] as happens for the rest of the elements. However, by inspecting their Fig. 5 we can see that the underabundance of [Ba/Fe] with respect to stars without detected giant planets takes place mainly at [Fe/H]\,$>$\,0\,dex meanwhile our sample does not show any difference at super-solar metallicities. The study by \citet{jofre_planets15} extends this Ba underabundance to the evolved stars population finding that planet hosts present by $\sim$\,0.11\,dex lower abundances on average. Although the difference they find is within the dispersion, it is systematic at [Fe/H]\,$>$\,--0.4\,dex. Using a small sample of 14 planet-host stars in the thin disk, \citet{mishenina16} also found this depletion in Ba. These authors also checked whether planet hosts in their sample were older than single stars since some works have reported a decline of [Ba/Fe] with age both in field stars \citep[e.g.][]{bensby07} and in clusters \citep[e.g.][]{dorazi09}, but they did find the same deficit of Ba at all ages. 

Stellar ages also play an important role in the GCE and contribute to the dispersion of abundances at similar metallicities \cite[e.g.][]{haywood13}. Indeed, the abundances of some elements present very clear trends with age. For example, the $\alpha$ elements show increasing [X/Fe] with age, i.e. at earlier times in the evolution of the Galaxy. These elements are mainly produced by massive stars, that were the first stars to die and contribute to the enrichment of the Galaxy, and have a negligible contribution to Fe. On the other hand, neutron-capture elements are produced in low-mass AGB stars which take longer to release their nucleosynthesis products to the Galaxy and then tend to be more abundant as the Galaxy evolves, i.e. for younger stars. The general [X/Fe] vs age trends for our sample can be seen in \citet{delgado17_nice}.

In order to check if age is the responsible for the depletion of Ba found in our planet hosts we show in Fig. \ref{Ba_age} the [Ba/Fe] ratios as a function of age for the same metallicity bins as in Fig. \ref{bins_XFe_Fe}. Stellar ages are derived from the PARAM v1.3 tool\footnote{http://stev.oapd.inaf.it/cgi-bin/param\_1.3} using the PARSEC theoretical isochrones from \citet{bressan12} and will be presented in a separate work (Delgado Mena et al. 2018, in preparation). The trend of decreasing Ba with age is quite well observed in the bins with --0.5\,$<$\,[Fe/H]\,$<$\,0.2\,dex. However, this trend tends to disappear at ages older than $\sim$4\,Gyr where the abundance ratios get mostly flat. In the panel with the whole sample we can see that high-mass planet hosts are well spread along different ages and abundances except in the age range 2-4 Gyr where they present clearly lower [Ba/Fe] on average than single stars. On the other hand, the low-mass planet hosts in our sample are older than 2 Gyr\footnote{Low-mass planets are easier to detect around non-active stars, older than 1-2\,Gyr, since for young and active stars the planetary signal is \textit{hidden} in the stellar noise.} and they seem to form like a lower envelope of [Ba/Fe] ratios along the age axis. Since low-mass planet hosts are older on average than single stars this could partially explain the Ba average underabundance. However, the RV detection bias towards older ages cannot be responsible for an abundance difference between planet host and non host stars that have ages of several Gyrs, since only very young stars are avoided in the search for planets. 

In Fig. \ref{Ba_age} we can see that when comparing the Neptunian hosts with single stars of similar age in each [Fe/H] bin they tend to show lower individual [Ba/Fe] (for [Fe/H]\,$<$\,0\,dex). Unfortunately, the number of low-mass planet hosts in each bin is not very high to draw a robust conclusion. The discovery of more low-mass planets in the future will give us the opportunity to confirm and understand the trend of Ba underabundance in such planet hosts. The fact that Neptunian hosts are older on average might partly explain the higher average [Zn/Fe] and lower average [Sr/Fe] abundances since those ratios increase and decrease for older stars, respectively, as shown in Figs. \ref{Zn_age} and \ref{Sr_age}. However, the older age cannot be the only responsible for such trends because then we should observe that the abundances of other neutron capture elements (showing a negative trend of abundance vs age) are also lower in planet hosts. This should be especially the case of [Y/Fe], which has a very clear decreasing trend with age \citep{nissen16,delgado17_nice,spina18}, but we don't see a systematic difference in Y abundances between stars with and without planets. On the other hand, systematic effects on abundances due to different stellar parameters can also produce biases in the comparison of different samples of stars.

\begin{figure}
\centering
\includegraphics[width=1.0\linewidth]{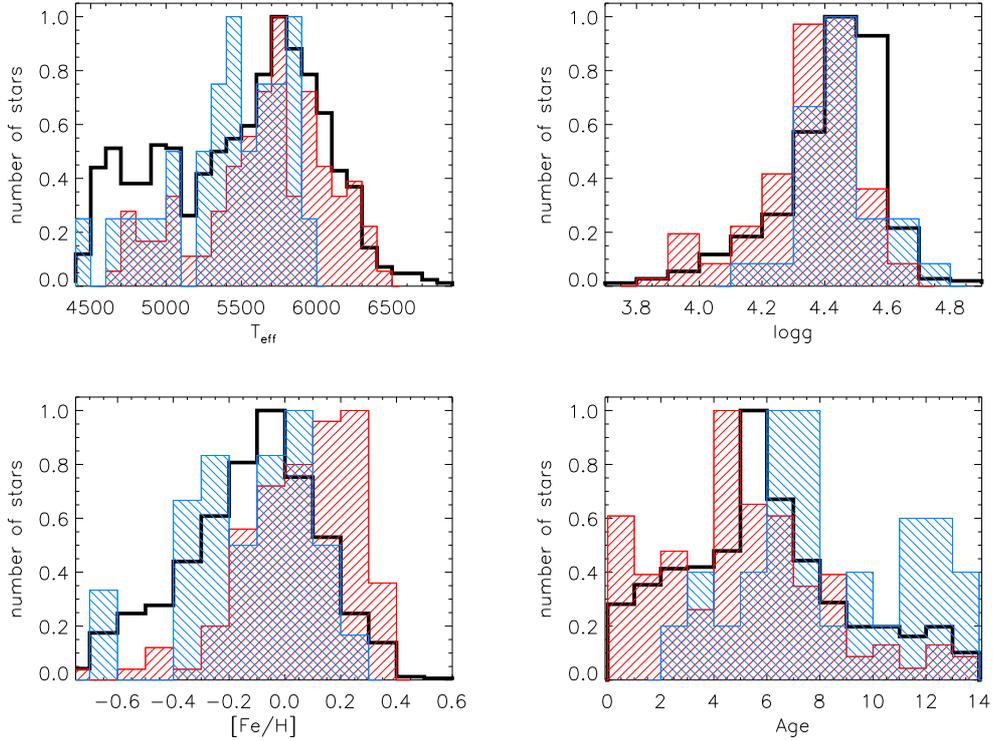}
\caption{Distribution of all stars in the sample (at [Fe/H]$>$--0.7). The thick black, red dashed and blue dashed histograms represent the single stars, Jupiter hosts and Neptunian hosts, respectively.}
\label{histogram_all}
\end{figure}

\subsection{Assessing the statistical significance of the differences in abundances}

In Fig. \ref{histogram_all} the distribution of stellar parameters for all our sample stars is shown where we can appreciate how Neptunian hosts are cooler and older on average than single stars. To evaluate the impact of different parameters we carry out several statistical tests considering the possible biases affecting our results. Such tests are detailed in the Appendix. We applied Kolmogorov-Smirnov tests to different samples of planet hosts and randomly selected single stars within a given range of parameters. Additionally, we perform a multivariable linear regression to account for the dependence of abundances on different parameters and establish whether there is an offset between the populations of stars with and without planets. We find that the difference initially detected for Zn is not seen anymore by applying a multivariable linear regression or a bootstrap resampling and is therefore a consequence of GCE as stated before. The situation for Sr, Ce and Zr is similar and the offsets found are not significant either. Only in the case of Ba we find that Neptunian hosts have an abundance 0.033\,dex lower than single stars with a significance of 3.1$\sigma$. This result should be considered with caution since by selecting stars in a narrower range of \teff\ the offset disappears although we note that the number of planet hosts is significantly reduced in such a sample.

\section{Conclusions}

In this work we present abundances of neutron-capture elements in the HARPS-GTO sample where there are 120 stars hosting high-mass planets and 29 stars hosting exclusively Neptunians and Super-Earths together with 910 stars without detected giant planets. We find that planet host stars (especially Neptunian-mass hosts) present higher abundances of Zn than single stars at [Fe/H]\,$<$\,--0.1\,dex. On the other hand, planet hosts show lower abundances of Ba, Sr, Ce and Zr than single stars. These differences are also more obvious at [Fe/H]\,$<$\,0\,dex. The fact that most of our planet hosts at low metallicities belong to the thick disk can in part explain the deviation found for Zn, since this element is enhanced in the thick disk, in a similar way as $\alpha$ elements. Moreover, a different distribution in the stellar parameters of each group of stars can account for at least a fraction of the differences in abundances. This is the case of Neptunian hosts, which are older on average, and thus are expected to have higher abundances of Zn on average and lower abundances of neutron-capture elements due to the effects of GCE.

We perform some statistical tests to understand whether the difference in abundances that we find is statistically significant and/or is an artifact of biases in the stellar parameters distributions of different samples. Such tests rule out the possibility that the presence of planets is affecting the abundances of these elements except for the case of Ba where a small underabundance in planet hosts is found.

Different theoretical studies on planet formation have shown that Fe, Si, Mg and O are the most important elements forming the bulk composition of rocky planets together with Ni and S. Therefore, we would not expect to observe strong differences between planet hosts and non-hosts for the elements analyzed here. Moreover, we could suppose that if an element is important for planet formation we would expect that planet hosts are enhanced in such element and not depleted, as observed for the case of Ba. Still, given that other authors have also found a difference in Ba abundances and our careful analysis also hints to a possible underabundance in low-mass planet hosts it would be interesting that this element is further studied by planetologists to see whether it can play any role on the formation of planets.\\

\textit{Acknowledgements}: E.D.M., V.A., P.F., N.C.S., S.G.S. and J. F. acknowledge the support from Funda\c{c}\~ao para a Ci\^encia e a Tecnologia (FCT) through national funds and from FEDER through COMPETE2020 by the following grants: UID/FIS/04434/2013\&POCI--01--0145-FEDER--007672, PTDC/FIS-AST/1526/2014\&POCI--01--0145-FEDER--016886, and PTDC/FIS-AST/7073/2014\&POCI-01-0145-FEDER-016880.
E.D.M., V.A., P.F., N.C.S., and S.G.S. also acknowledge the support from FCT through Investigador FCT contracts IF/00849/2015/CP1273/CT0003, IF/00650/2015/CP1273/CT0001, IF/01037/2013/CP1191/CT0001, IF/00169/2012/CP0150/CT0002, and IF/00028/2014/CP1215/CT0002. J.F. acknowledges support by the fellowships SFRH/BD/93848/2013 funded by FCT (Portugal) and POPH/FSE (EC). J.I.G.H. acknowledges financial support from the Spanish Ministry of Economy and Competitiveness (MINECO) under the 2013 Ram\'on y Cajal program MINECO RYC--2013--14875, and the Spanish ministry project MINECO AYA2014--56359-P.\\

This research has made use of the SIMBAD database operated at CDS, Strasbourg (France) and the Encyclopaedia of Extrasolar Planets.

\bibliographystyle{aa}

\bibliography{edm_bibliography} 

\section{Appendix: Statistical tests}
 
As explained before, [X/Fe] ratios present a dependence on metallicity and age. Moreover, abundances tend to slightly vary with \teff\ (see DM17) and \logg\ (for the ionized species). Therefore, we performed different bootstrap resampling tests with selected subsamples controlling the different stellar parameters. In each test we choose the planet hosts we want to compare, and then we select mock samples of random non-repeated single stars with the same number of objects as the planet hosts sample. The selection of mock samples is repeated 10000 times and in each iteration the abundances of both samples is compared. Moreover, in some tests we also apply a Kolmogorov-Smirnov (K-S) to compare the distribution of different parameters between the planet hosts sample and the mock samples of single stars. If the p-value of such tests is higher than 0.05 we consider that the K-S test fails to reject the hypothesis that both samples belong to the same parent population. Now we detail the specific tests we carried out:

\begin{enumerate}
 \item A first simple test is to select solar \teff\ stars where the errors on abundances are minimal. Therefore, we selected Neptunian hosts with  with \teff$_{\odot}$\,$\pm$\,300\,K and \logg$_{\odot}$\,$\pm$\,0.2\,dex\footnote{Our adopted solar parameters are \teff$_{\odot}$\,=\,5777\,K, \logg$_{\odot}$\,=\,4.43\,dex.}. We also restricted this subsample to stars older than 7\,Gyr (to avoid a comparison with younger single stars, see Fig. \ref{histogram_all}) having in total 10 Neptunian hosts. For comparison, we randomly selected 10 single stars within this range of parameters and with --0.4$<$[Fe/H]$<$0.2, which is the metallicity range of Neptunian hosts. We find that while the distribution of average values of [Fe/H] and age in the mock samples is centered around the average values for Neptunian hosts, the average of [Ba/Fe] abundances of Neptunian hosts is lower than for single stars in $\sim$\,95\% of the iterations (see upper panels of Fig. \ref{test1}). On the other hand, for Zn and Sr ratios the difference found in previous subsections disappears and the average values of mock subsamples are centered around the average value for Neptunian hosts. An analogous test can be done to compare single stars to Jupiter hosts, using the same range in parameters except for metallicity which lays between --0.6 and 0.4\,dex, and for age, which is not needed to be constrained since the distribution of ages for both groups is similar. In this subsample we have 51 planet hosts. In this case we also find that the distributions of Zn and Sr are very similar between observed and mock data in each iteration (see lower panels of Fig. \ref{test1}). However, the average of Ba abundances in single stars is always higher than for Jupiter hosts despite being slightly older on average and significantly less metallic, two characteristics that favor a lower Ba content. From this simple test it is clear that the previously difference in Zn  and Sr abundances is just an artifact of planet hosts having a different distribution of stellar parameters. The weak point of this kind of test is that to ensure a similar distribution of parameters between both samples we end up with a very small sample of Neptunian hosts to check.\\

\item Thus, we performed a second bootstrap resampling in which we consider all the stars hosting Neptunians for which we could determine ages (28). In each iteration we randomly draw 28 different comparisons stars that match the stellar parameters of each planet host one by one, i.e. we find a couple for each planet host. We consider a maximum difference of 100\,K in \teff, 0.10\,dex in [Fe/H], 0.2\,dex in \logg, 0.2\,M$_{\odot}$ in mass and 2\,Gyr in age. We find that the average of Ba and Sr abundances for Neptunian hosts is lower than for the mock samples of single stars in 100\% and 99.8\% of the iterations whereas for Zn the average values of single stars are centered around the value for Neptunian hosts (in 44\% of the iterations we find higher abundances in Neptunian hosts). In addition, for each iteration we run a one-side K-S test to compare the distribution of each of the different stellar parameters between the planet hosts sample and the mock samples. The K-S tests fails to reject the hypothesis that the stellar parameters of both samples are drawn from the same population with p-values always higher than 0.4. The p-values for the distribution of Zn, Sr and Ba show lower values but in very few of the 10000 iterations they have a value lower than 0.05, a requirement to state that both samples do not come to the same parent population. 
When we apply this same test to the Jupiter hosts (115 stars with derived ages) compared to single stars we find that [Zn/Fe] and [Ba/Fe] ratios are higher and lower in the planet hosts sample in 38\% and 62\% of the iterations, respectively, when compared to the mock samples of single stars. Curiously, the behaviour for Sr is inverted and Jupiter hosts present a lower average abundance only in 5\% of the iterations. The K-S tests applied to each iteration show that the distribution of stellar parameters and abundance ratios for the three elements in both groups can be drawn from the same population with p-values always higher than 0.05. This test has the disadvantage of selecting very similar mock samples (or even repeated) after few iterations since we are applying strong constrains on the similarity of each couple.\\

\begin{figure}
\centering
\includegraphics[width=0.8\linewidth]{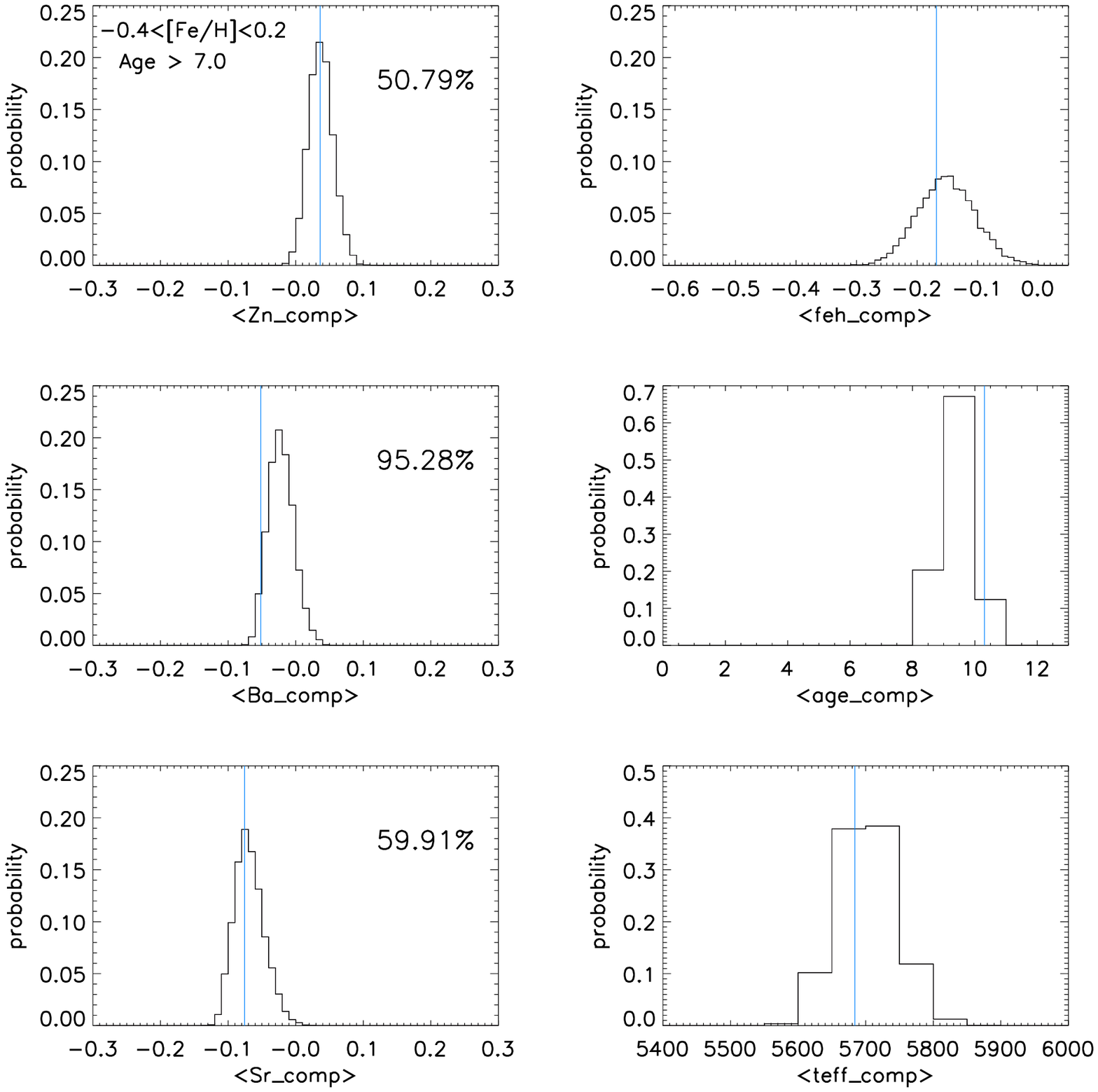}
\includegraphics[width=0.8\linewidth]{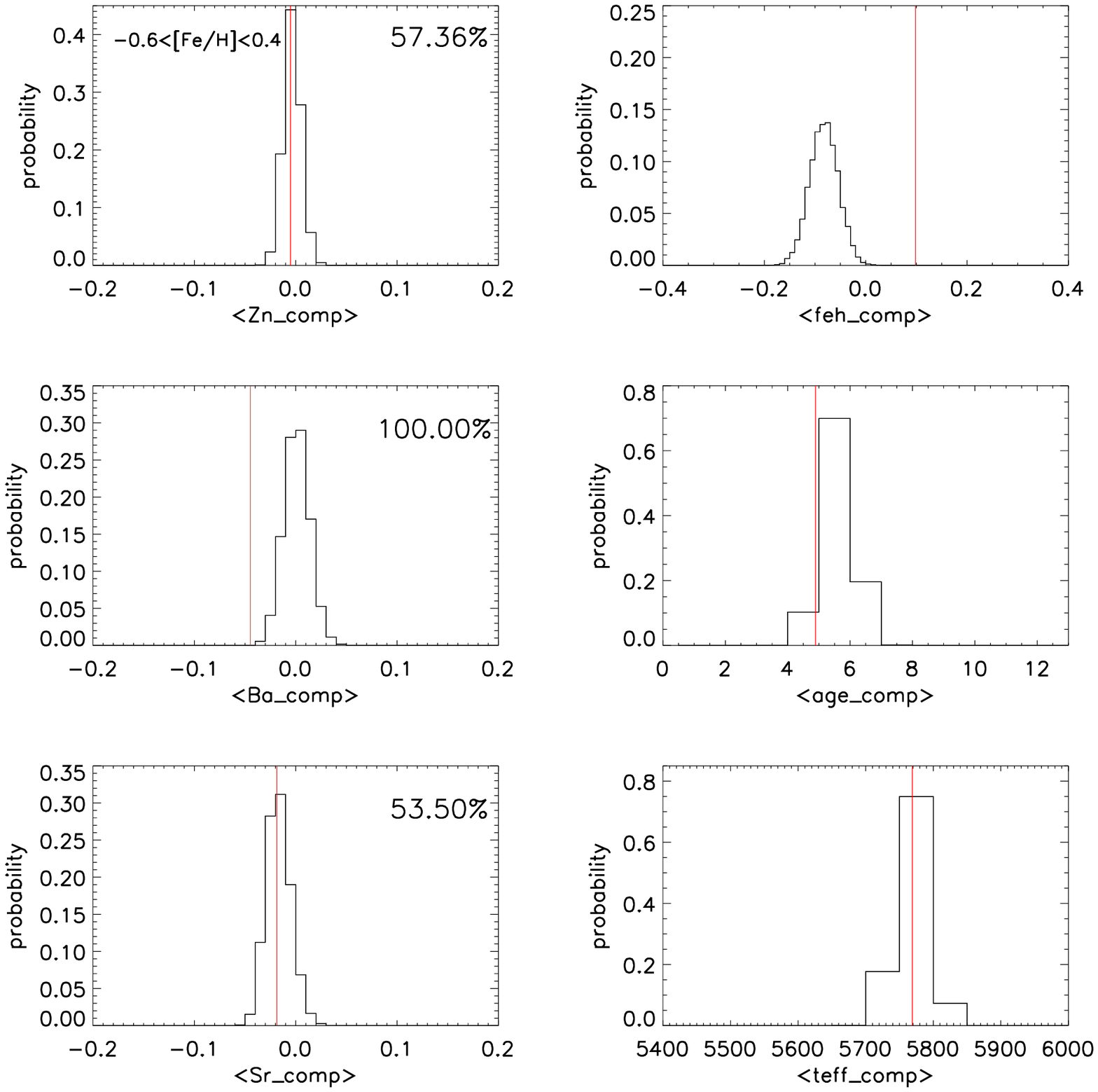}
\caption{Distribution of average values for different parameters in each of the mock samples randomly selected. The blue and red lines mark the average values of the Neptunian and Jupiter samples for the first test. The percentage shows the number of times in which the average abundance of Ba and Sr is lower in planet hosts than in single stars and the average abundance of Zn is higher in planet hosts than in single stars.}
\label{test1}
\end{figure}

\item We carried out a third bootstrap resampling selecting mock samples of single stars with no constrain on the stellar parameters. In each iteration a one-side K-S test is performed to compare the stellar parameters of the mock samples with the planet hosts sample. If the p-values for the comparison of all the stellar parameters are above 0.05 we consider that such mock sample has a similar distribution of parameters as the planet hosts sample so we keep it and apply the K-S test to the abundance ratios. After 10000 iterations we find that only in 7\% and 12\% of the mock samples the p-values of [Zn/Fe] and [Sr/Fe], respectively, are lower than 0.05 whereas in 52\% of the samples the p-values for the comparison of [Ba/Fe] are lower than 0.05. Although this is not a significant result it shows that for many of the mock samples there seems to be a difference in Ba abundances that is interesting to further explore. Again, this test has the disadvantage of selecting repeated samples of stars despite running a high number of iterations.

\end{enumerate}

Therefore, in order to handle the partial collinearity of the variables and to correct for the dependence on each of these on the abundances, we perform a multivariable linear regression which allows to use the full sample of stars. The regression is applied to the planet host sample (Jupiters or Neptunians in a separated way and together) and the comparison sample as explained in \citet{figueira14}. To apply this test we assume that there is a linear dependence of the abundance ratio [X/Fe] on stellar parameters (\teff, [Fe/H], \logg, age) on first approximation and thus we want to fit an equation like this:

\vskip-1.5em
\begin{eqnarray}\label{eq1}
\mathrm{[X/Fe]} & = & int. \,+\, \beta_1 \rm T_{\rm eff} \,+\, \beta_2 \rm [Fe/H] \,+\, \beta_3 \mathrm{log}\,g \,+ \\ \nonumber
                       && + \, \beta_4 \mathrm{Age} \,+\, M\,\times\,\mathrm{offset}
\end{eqnarray}

in which \textit{int.} represents the intercept value, which is the value of the dependent variable [X/Fe] when all the independent variables are zero, and $\beta_{1-4}$ are the coefficients of the linear regression associated with each independent variable.

\begin{table*}
\centering
\begin{tabular}{cccccccccc} \hline\hline
sample & number & {\it int.} &$\beta_1$ & $\beta_2$  & $\beta_3$  & $\beta_4$   & offset & significance \\ 
\hline
Zn jupiters &112PH \& 841CS &   8.5e-2  & -5.1e-6 &   2.3e-2  & -3.6e-2 &   1.6e-2 &   0.008  &  1.2 \\
Zn neptunes & 28PH \& 841CS &   1.3e-1  & -7.2e-6 &   1.4e-2  & -4.2e-2 &   1.5e-2 &   0.002  &  0.2 \\
Zn all planets  &140PH \& 841CS &   7.8e-2  & -5.3e-6 &   2.0e-2  & -3.4e-2 &   1.6e-2 &   0.007  &  1.2 \\
\hline
Ba jupiters &112PH \& 841CS &  -1.2e+0  &  4.7e-5 &  -1.4e-1  &  2.2e-1 &  -9.0e-3 &  -0.012  &  1.9 \\
Ba neptunes & 28PH \& 841CS &  -1.2e+0  &  4.8e-5 &  -1.2e-1  &  2.2e-1 &  -8.5e-3 &  -0.033  &  3.1 \\
Ba all planets  &140PH \& 841CS &  -1.2e+0  &  4.7e-5 &  -1.4e-1  &  2.1e-1 &  -9.0e-3 &  -0.016  &  3.0 \\
\hline
Ce jupiters &112PH \& 841CS &  -4.0e+0  &  1.2e-4 &  -4.6e-2  &  7.6e-1 &   1.4e-3 &  -0.011  &  1.5 \\
Ce neptunes & 28PH \& 841CS &  -4.3e+0  &  1.4e-4 &  -4.2e-2  &  8.0e-1 &   8.8e-4 &  -0.055  &  2.5 \\
Ce all planets  &140PH \& 841CS &  -4.0e+0  &  1.2e-4 &  -4.0e-2  &  7.5e-1 &   1.2e-3 &  -0.022  &  2.5 \\
\hline
Sr jupiters &112PH \& 841CS &   2.5e-2  & -9.0e-5 &  -1.2e-1  &  1.3e-1 &  -1.6e-2 &   0.004  &  0.3 \\
Sr neptunes & 28PH \& 841CS &   4.7e-2  & -8.9e-5 &  -1.1e-1  &  1.3e-1 &  -1.5e-2 &  -0.024  &  1.4 \\
Sr all planets  &140PH \& 841CS &   3.8e-2  & -8.9e-5 &  -1.2e-1  &  1.3e-1 &  -1.6e-2 &  -0.003  &  0.3 \\
\hline
Zr jupiters &112PH \& 841CS &  -2.9e+0  &  3.7e-5 &  -1.8e-1  &  6.1e-1 &  -3.5e-3 &  -0.012  &  1.5 \\
Zr neptunes & 28PH \& 841CS &  -3.0e+0  &  4.1e-5 &  -1.7e-1  &  6.4e-1 &  -3.3e-3 &  -0.048  &  2.2 \\
Zr all planets  &140PH \& 841CS &  -2.8e+0  &  3.6e-5 &  -1.7e-1  &  6.0e-1 &  -3.8e-3 &  -0.020  &  2.2 \\
\hline 
\end{tabular}
\caption{Parameters for each coefficient as resulting from multivariable linear regression analysis in different tests where we restrict the samples (PH are planet hosts and CS are the comparison stars) to [Fe/H]\,$>$\,--0.7 and \logg$>$\,4. The offset is only included in the fit for planet hosts since the offset for comparison stars is by definition 0. The last column reflects the significance of the offset found between both samples.}
\label{test}
\end{table*}

The regression is done simultaneously for both data sets (planet hosts and single stars) which have the same coefficients $\beta_i$ but allowing for an offset for the planet host sample (so by definition \textit{M}\,=\,0 for single stars and \textit{M}\,=\,1 for planet hosts). This offset represents the difference between both data sets and cannot be caused by different stellar parameters. This kind of test permits us to use the full sample of stars without restricting the range of stellar parameters as happened in previous tests. The only constrain in the sample is made to remove single stars with [Fe/H]\,$<$\,--0.7\,dex (because there are no planet hosts below that metallicity) and the few subgiants (\logg$<$\,4) we have in our sample. In Table \ref{test} we show the results for the regression. The significance of the offset was estimated by bootstrapping the abundance ratios with a Gaussian distribution of the errors around the corresponding uncertainty on each abundance ratio. The regression was then repeated during 10000 times using the bootstrapped values. As an additional check, for the cases where we find a big or significant difference in abundance, we also apply the test to a restricted sample with \teff$_{\odot}$\,$\pm$\,500\,K, where the errors on abundances are lower. We decided to apply the regression also to Zr and Ce, because these abundances have in general larger errors and with the previous tests we were not taking them into account. We can see how Ba, Sr and Zr show a negative correlation with age ($\beta_4$) and [Fe/H] ($\beta_2$) in the multivariable regression while Zn shows a positive correlation with these two parameters. On the other hand, there is a positive correlation with \logg\ ($\beta_3$) for ionized elements (all except Zn).

All the tests show that the offsets are small, in many cases lower than the average errors on abundances, and only in few cases this offset has a significance above 3$\sigma$. For example, [Ba/Fe] is --0.033\,dex lower in Neptunians than in single stars at the level of 3.1$\sigma$. However, if we check for this difference in the narrower range of 'solar' \teff\ stars the offset goes below 0.01\,dex and is not significant. For the case of [Sr/Fe] we can find an offset of --0.023\,dex in Neptunians which is not significant though (1.4$\sigma$). For Zn the differences are very small both for Jupiters or Neptunians. The largest offsets are found for Zr and Ce (0.048 and 0.055\,dex lower on Neptunians when compared to single stars) although they are at the level of 2$\sigma$. We note that we have applied the same test by comparing the full sample of planet hosts (regardless of their masses) with single stars and the outcomes are similar. In summary, with these results we cannot claim for any statistically significant difference between the abundances of planet hosts (either Neptunians or Jupiters) and single stars except for maybe the case of Ba which would be interesting to review in the future.\\

\end{document}